\documentclass[runningheads]{svmult}

\usepackage{makeidx}   
\usepackage{graphicx}  
\usepackage{subeqnar}  
\usepackage{multicol}  
\usepackage{physprbb}  
\makeindex             

\begin{document}
\title*{Extragalactic Planetary Nebulae: Observational Challenges \& Future Prospects}
\toctitle{Extragalactic Planetary Nebulae:
\protect\newline Observational challenges and Future Prospects}

\titlerunning{EPN: Challenges \& Prospects}
%
\author{Quentin A Parker\inst{1}
and Richard Shaw\inst{2}}
\authorrunning{Parker \& Shaw}
%
%
\institute{Macquarie University/Anglo-Australian Observatory, Sydney, Australia,
qap@ics.mq.edu.au
\and NOAO, USA, shaw@noao.edu}

\maketitle              

\section{Introduction}

The study of extragalactic planetary nebulae (EPN) is a rapidly expanding field. The advent of
powerful new instrumentation such as the PN spectrograph (e.g. Douglas et al. these proceedings) 
has led to an avalanche of new EPN discoveries both within and between galaxies. We now have 
thousands of EPN detections in a heterogeneous selection of nearby galaxies and their local 
environments, dwarfing the combined galactic detection efforts of the last century. This alone 
brings its own problems of nomenclature and cataloging (Parker \& Acker, these proceedings).

Key scientific motivations driving this rapid growth in EPN research and discovery have been the
use of the PNLF as a standard candle (e.g. Ciardullo, Girardi these proceedings),
the use of EPN as dynamical tracers of their host galaxies and of dark matter 
(e.g. Napolitano, these proceedings),
and as probes of Galactic evolution (e.g. Liu, Richer, these processings).
This is coupled with the basic utility of PN as laboratories of nebula physics and the consequent
comparison with theory where population differences, abundance variations and star formation
history within and between stellar systems informs both stellar and galactic evolution.

Here we pose some of the burning questions, discuss some of the observational challenges and 
outline some of the future prospects of this exciting, relatively new, research area as we strive 
to go fainter, image finer, see further and 
survey faster than ever before and over a wider wavelength regime.

\section{Some selected questions}
The ability to not only detect but to begin to study  PN in external galaxies permits us
to pose some new questions such as:
What is the origin of the PNLF, how far can we push it and does it break down? 
How does the nature of a host galaxy affect SFH, 
chemical evolution and the PN population? Does it play a role in nebular morphology?
Can multi-galaxy studies provide clues to the progenitor/core mass relation for medium 
to low mass stars? What is the nature and extent of the intra-cluster stellar population 
and what can it tell us of the merger histories via detection of evidential streams? Finally what
is the role of dust in PN formation and evolution?

\section{The role of technology}
The available technology and instrumentation
both drives and limits what we can observationally achieve at any given epoch and in 
any wavelength domain. We need to develop the most viable strategies for tackling the above
questions and push to influence the development of the facilities necessary to execute these 
strategies effectively. The PN spectrograph is a rare example of an instrument
specifically designed to detect PN. Do we want more of these? 
Usually we have to make use of instruments that may not be
ideally suited to our needs leading to compromise or adaptation of the science mission. We need to
ask if the currently available technology is driving our research directions more than we are
driving the technology. It behoves our community to take an active role in influencing 
the design and capability of the next generation of instrumentation to better fit requirements. 

Most current EPN detection techniques perform on-band off-band imaging for a single (usually the
[O III] 5007\AA) line. Contamination is a serious issue. The ability to efficiently image several nebula
lines in the rest frame of a galaxy over a wide field of view would be a significant step forward.
Larger/multi aperture optical telescopes of 8m+ now outnumber the 4m class which are supplemented
by new multi-wavelength capabilities from earth and space in the UV/IR/MIR and radio regimes.
Higher resolution capabilities coupled with more efficient/powerful detectors, spectrographs, 
interferometers, Fibre/IFU/multi-slit technologies, narrow band capabilities and
innovative techniques (e.g. CCD nod and shuffle and tunable filters) offers significant
opportunities to advance EPN research. 

Tunable filters offer much promise. They can rapidly image a series of diagnostic 
nebular lines over a wide field at a given redshift and can be used to scan through an entire 
depth of a cluster. The Taurus Tunable Filter on the AAT was a highly effective tool in this regard
but although a powerful and unique capability it has now been de-commissioned. We need to ensure
that the best facilities suited to our needs are maintained and enhanced.

Examples of ground based instrumentation advances on 8m telescopes 
include use of optical 
fibres in MOS and IFU systems such as FLAMES/GIRAFFE (ESO VLT) and FMOS/Echindna (Subaru); 
enhanced IR imaging capabilities such as VISIR (VLT) and advances in active and adaptive optics systems 
(including MCAO, especially in the NIR). New space-based facilities such as Spitzer 
offer new multi-wavelength sensitivities while
planned future facilities such as SKA, OWL and the JWST will offer unique capabilities.
We need to plan now to ensure that we achieve access to such facilities
and influence their development.

\section{The observational \& theoretical challenges}
An utlimate aim is to obtain for EPN the observational detail we obtain for PN in our own 
Galaxy. Hence, aside from mere EPN detection, we would like to:\\
$\bullet$ detect/study their central stars (e.g. Villaver, these proceedings); \\
$\bullet$ detect/study their outer halos (now just possible for the LMC 
- Reid \& Parker, these proceedings); \\
$\bullet$ detect highly evolved examples and determine PN morphologies 
and nebular abundances. Currently only  possible to some degree for 
Magellanic Cloud PN via the HST (e.g. Stanghellini \& Shaw, these proceedings).\\ 
$\bullet$ obtain high S/N, decent resolution spectra permitting EPN nebular parameters to be 
determined across a wide wavelength range coupled with accurate radial and expansion velocities.

Faster computers 
and more powerful, precise and sophisticated n-body simulations and photoionisation code coupled to
significant new observational EPN data would enable us to fine tune theory/models of diverse 
PN populations in widely different galaxy environments.

\subsection{Going fainter \& seeing further}
If we simply probe fainter via a combination of light gathering power (larger telescopes), 
system efficiency gains (e.g. by AO), sky/background suppression (e.g. narrow-band filters) 
and exposure times, we can push the PNLF to fainter limits in external galaxies at greater 
distances as the most luminous PN in more distant systems move into the detection threshold. 
Spectroscopically we can obtain higher S/N observations of more useful nebular diagnostic lines 
in our own galaxy, LMC and local group and perhaps begin to sample basic lines in more distant 
galaxies. Abundance determinations in systems $<10$Mpc would become feasible together with the
detection of halos, additional morphological features and lower surface brightness PN.

\subsection{Obtaining finer resolution}
Improved resolution imaging and spectroscopy (across a wider wavelength domain) will provide the
ability to discriminate finer morphological detail in EPN (currently only possible in the LMC via
the HST) and permit the study of reaction interfaces between EPN and their local ISM. 
Measurements of accurate expansion velocities and systemic velocities in external 
systems would permit improved kinematical studies and the removal of contaminants in spectra 
that were overlapping in lower resolution systems.
 
\subsection{Surveying faster}
We are truly entering an era of the data avalanche in astronomy as a consequence of
large survey facilities, focal plane arrays, advanced MOS systems and other significant efficiency
gains in coverage, resolution and sensitivity. The current
AVO/Astrogrid initiatives to federate, incorporate and manage the massive petabyte catalogues now
being created in astronomy have been set-up to meet this challenge. These are
coupled with vast storage and processing capabilities as technology and Moores law allows.
We need to be able to handle/understand cross-correlate and disseminate our own EPN results 
in a timely fashion and make best use of these new AVO capabilities.

\section{Problems \& issues arising from this meeting}

The most significant problem is the extreme faintness of EPN even with line-imaging. This is coupled
with poor resolution (EPN are unresolved in all but the 
closest galaxy neighbours with current technology). Another issue is the contamination of EPN
samples when only single lines are imaged. However
soon we will have tens of thousands of individual EPN crying out for further study and careful
cataloguing. It is also 
clear that we badly need better and indeed any spectra of PN candidates in the intracluster 
medium. We also urgently need better simulations to leverage observations.
We want to determine multi-dimensional PNLFs and directly detect EPN central stars and 
determine individual spectral and photometric properties. Such problems represent a considerable
observational challenge. Only strategic planning, collaboration and technological
developments are likely to be able to address these problems effectively.

\section{Collaborations \& future prospects}
A community speaking with one voice can be heard more effectively and a community co-operating to
achieve a common goal is more likely to suceed. The prohibitive cost of new instrumentation and the
plans for 50m+ telescopes necessitates large multi-national collaborations. It is vitally
important that our community has an influence when key instrumentation capabilities are chosen.
Furthermore it
may prove friutful to inaugurate several key large-scale community wide collaborations to develop
effective exploitation strategies to use these powerful new facilities and to tackle the really big
issues such as indicated below.

While it will be some time before we can effectively study individual EPN in all but the closest
galactic neighbours, currently available capabilities mean that there are realistic prospects of
advancement is several key areas. 
Should we aim for a complete EPN census of Local Group galaxies, complete to $\sim5$ mag below the 
L(5007) peak and a census of intra-cluster EPNs to depths of several Mpc? This will be invaluable
in understanding the origins of the PNLF and what occurs in late type populations and how EPN
inform host galaxy SFH. Can we perform morphological studies of complete samples of EPN in at 
least 2 local group galaxies, e.g. via HST/STIS or MCAO on 8m telescopes?
Should we obtain a census of intra-cluster EPN in Virgo/Fornax together with velocity estimates? 
This will impact on our understanding of galaxy dynamics, the IC stellar population and the
influence of dark matter.\\

\noindent{\bf Acknowledgements}\\
The authors are grateful for the positive feedback from the participants during the course of the
workshop and during the scheduled discussion.

%
%

%

\end{document}